\documentclass[aps,showpacs,prl,preprint]{revtex4-1}
\usepackage{epsfig}
\usepackage{amsmath}
\usepackage{hyperref}

\begin{document}
\title{Electron rescattering at metal nanotips induced by ultrashort laser pulses}
\author{Georg Wachter}
\author{Christoph Lemell}
\author{Joachim Burgd\"orfer}
\affiliation{Institute for Theoretical Physics, Vienna University of Technology, Wiedner Hauptstr.\ 8-10, A-1040 Vienna, Austria, EU}
\author{Markus Schenk}
\author{Michael Kr{\"u}ger}
\author{Peter Hommelhoff}
\affiliation{Max Planck Institute of Quantum Optics, Hans-Kopfermann-Str.\ 1, D-85748 Garching, Germany, EU}

\date{\today}

\begin{abstract}
We report on the first investigation of plateau and cut-off structures in photoelectron spectra from nano-scale metal tips interacting with few-cycle near-infrared laser pulses. These hallmarks of electron rescattering, well-known from atom-laser interaction in the strong-field regime, appear at remarkably low laser intensities with nominal Keldysh parameters of the order of $\gtrsim 10$. Quantum and quasi-classical simulations reveal that a large field enhancement near the tip and the increased backscattering probability at a solid-state target play a key role. Plateau electrons are by an order of magnitude more abundant than in comparable atomic spectra, reflecting the high density of target atoms at the surface. The position of the cut-off serves as an in-situ probe for the locally enhanced electric field at the tip apex.
\end{abstract}
\pacs{79.20.Ws, 32.80.Rm, 79.60.-i, 79.70.+q}
\maketitle

Since the early days of quantum physics, photoemission from solid surfaces has played a key role in both probing structure and dynamics of surfaces and in exploring conceptual aspects of light-matter interaction. With the availability of intense femtosecond laser pulses, a novel regime beyond one-photon absorption (or linear response) spectroscopy such as X-ray photoelectron spectroscopy (XPS) has opened up. Resonant multi-photon processes allow to probe collective excitations and localized states in the band gap \cite{Petek1997}. With increasing intensity non-linear processes such as above-threshold photoemission (ATP) peaks appear in the spectra, which are spaced by multiples of $\hbar\omega$ corresponding to photoabsorption of a large number of photons well in excess of the ionization threshold \cite{Luan1989,Fann1991,Banfi2005,Schenk2010}.

In the strong-field regime delimited by small Keldysh parameters \cite{Keldysh65}, $\gamma=\sqrt{W/2U_\mathrm{p}}\lesssim 1$ with $U_\mathrm{p}=F_0^2/4\omega^2$ the ponderomotive energy ($F_0$: peak amplitude of field, $\omega$: laser frequency, $W$: work function) new features are expected to appear: a plateau in the photoemission spectrum that extends up to a cut-off at energies of $10 U_\mathrm{p}$. These signatures of strong-field physics have been observed for atoms and molecules \cite{Paulus1994_PRL} and with dielectric nanospheres \cite{Zherebtsov2011}. Their description relies on classical rather than on quantum interaction processes of the radiation field with the target: the electron emitted near the field maximum during a laser cycle is driven back to the core with energies of $\sim 2 U_\mathrm{p}$. Upon rescattering it gains additional energy of up to $10 U_\mathrm{p}$ \cite{Paulus1994}. For very short few-cycle pulses the ATP spectrum becomes sensitive to the carrier-envelope phase $\phi_{\mathrm{CEP}}$ of the laser pulse defined by $F(t)=F_0\cdot f(t)\cdot\cos(\omega t+\phi_{\mathrm{CEP}})$ with an envelope function $f(t)$.

Exploration of strong-field phenomena at solid surfaces has remained elusive. The range of usable intensities is limited by the threshold for surface damage. CEP dependencies have been predicted \cite{Lemell2003} and observed, however, only at a very low contrast level \cite{Apolonski2004}, most likely because the surface projected spot size at grazing incidence exceeds by far the wavelength $\lambda$ of the driving pulse. Interaction of few-cycle laser pulses with nano-scale metal tips offers the opportunity to explore strong-field effects in the condensed phase bypassing some of the difficulties encountered for extended surfaces. Such hybrid systems combine very small emission areas with linear dimensions small compared to $\lambda$ with the effects of the broken inversion symmetry of a solid surface. Moreover, field enhancement strongly localized at the tip apex affords the opportunity to observe strong-field physics without surpassing the damage threshold.

In this letter we investigate, both experimentally and theoretically, the ATP spectrum emitted from a tungsten nanotip irradiated by a few-cycle near-infrared (NIR) laser pulse at moderate intensities $I_0\sim 10^{11}$ W/cm$^2$, corresponding to $\gamma_0=\sqrt{W/2U_p} \gtrsim 10$ deep in the multiphoton regime. Remarkably, we observe unambiguously the strong-field signatures of a plateau and the cut-off associated with electron rescattering at the tip. Simulations of the local electromagnetic field near the tip and of the driven electronic dynamics, the latter both classically and quantum mechanically, show that the nano-scale confined dielectric response leads to a strong enhancement of the electromagnetic driving field by a factor $\lesssim 10$, thereby reducing the nominal Keldysh parameter $\gamma_0$ to an effective value of $\gamma_{\mathrm{eff}}\approx 2$. Furthermore, rescattering is strongly enhanced compared to the atomic analogue in the gas phase due to the high solid-state target density.

The experimental setup is described in detail in \cite{Schenk2010,Krueger2011}. Briefly, ultrashort ($\sim 6.5$ fs, full width at half maximum of the intensity) linearly polarized few-cycle laser pulses with a central wavelength of $\lambda=800$ nm (photon energy $\sim 1.55$ eV) are focused ($1/e^2$ spot radius $\sim 1.8\,\mu$m) on a sharp tungsten tip with a tip radius of about 6 nm. The polarization direction of the laser pulse coincides with the surface normal at the tip apex, $\phi_{\mathrm{CEP}}$ varies randomly from pulse to pulse. The pulses are derived from a Ti:sapphire oscillator with 80-MHz repetition rate, which allows us to obtain significant statistics even for very low electron emission rates (electron yield $\lesssim 1$ e$^-$ per pulse). The energy of electrons emitted from the surface is measured with a retarding field spectrometer with an effective resolution of about 0.5 eV (including smoothing of the spectra) over the energy range observed. Additionally, we observe the spatial emission characteristics by an imaging microchannel plate detector (field-emission and field-ion microscope setup), from which we determine the central emission region at the tip apex to coincide with the crystallographic W(310)-orientation. Spectra were taken in the presence of an extraction voltage of 50 V, resulting in an effective dc electric field near the tip of $F_{\mathrm{dc}}= 0.7$ GV/m  $\approx 1.4\cdot 10^{-3}$ a.u.\ when dc field enhancement due to the sharp structure is taken into account. This choice of $F_\mathrm{dc}$ assures the appearance of the plateau with a sufficient count rate also at small laser intensities.

\begin{figure}
\centerline{\epsfig{file=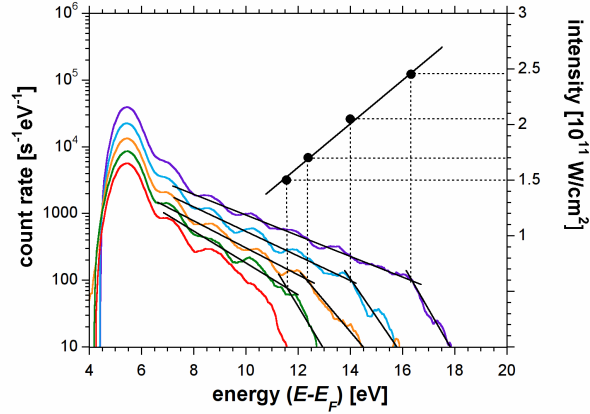,width=8cm}}
\caption{(Color online) Energy spectra of electrons emitted from a tungsten nanotip (radius about 6 nm) interacting with moderately intense laser pulses. Plateau and cut-off regions have been highlighted with lines. The energy axis is referenced to the Fermi level. The electronic kinetic energy is reduced by the workfunction of the tip ($W_\mathrm{W(310)}=4.35$ eV). The black dots show the cut-off energies $E_\mathrm{cut}$ determined from the intersection points of the lines as function of intensity. The solid black line fit shows the linear increase of $E_\mathrm{cut}$ with increasing intensity.}
\label{fig1}
\end{figure}
Energy spectra as a function of the laser intensity between $I_0=1.3\cdot 10^{11}$ and $2.4\cdot 10^{11}$ W/cm$^2$ display clear strong-field features, in particular a plateau followed by a sharp cut-off (Fig.\ \ref{fig1}). This is at first glance surprising in view of the large nominal Keldysh parameter ($\gamma_0\gtrsim 10$) or modest field amplitude ($F_0\lesssim 2.5\cdot 10^{-3}$ a.u.) involved. As expected for atomic spectra, the cut-off energy increases linearly with intensity (Fig.\ \ref{fig1}). Two unusual features are noteworthy: the cut-off kinetic energy of about 12 eV (purple, top spectrum) would require a strongly enhanced field of $F_{\mathrm{eff}}\approx 0.025$ a.u.\ to be compatible with the estimate of $10 U_\mathrm{p}$. Second, the plateau region relative to the dominating direct peak is much stronger in yield than in typical atomic spectra. In the atomic case the relative height is typically $\sim 10^{-3}-10^{-2}$ at comparable intensities \cite{Schaetzel2006PhD,Gazibegovic-Busuladzic2010}, whereas here it is about $0.05$.

\begin{figure}
\centerline{\epsfig{file=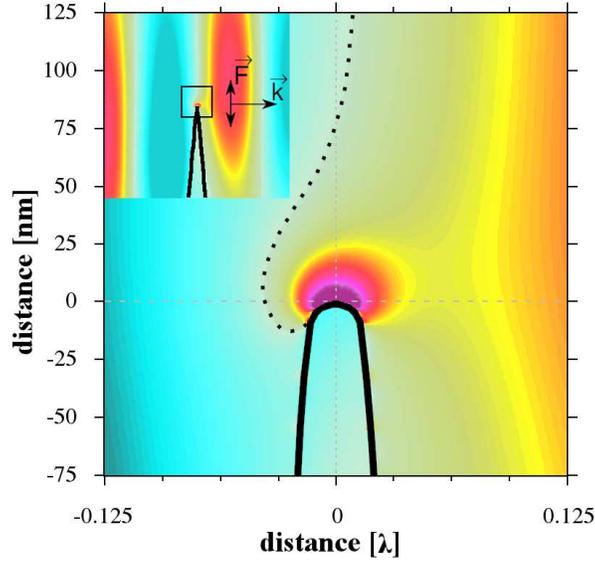,width=8cm}}
\caption{(Color online) Close-up of tip apex (200 $\times$ 200 nm) (inset: overview with 1200 $\times$ 1200 nm): cut through the field distribution along the direction of the tip axis with a tip radius of 10 nm. To visualize the phase shift the distance in $\vec k$ direction is given in units of $\lambda$. The presence of the tip visibly distorts the electric field. While the exciting field has a zero crossing near $x=0$ in $\vec k$ direction (black dotted line) the electric field at the apex is near its maximum (purple). Here, the field is enhanced by about a factor 5, $\Delta \phi_\mathrm{CEP}\approx 0.45\pi$.}
\label{fig2}
\end{figure}

In order to uncover the origin of these strong-field effects we have performed both quasi-classical and quantum simulations. We solve Maxwell's equations for the time-dependent field in the vicinity of the metallic tip using the finite differences time domain (FDTD) method \cite{Kunz1993Finite,taflove:2005}. The dielectric response of the tip is described by a discontinuous dielectric function 
\begin{equation}
\varepsilon (\omega,\vec r)=\left\{
\begin{array}{c@{\quad\ldots\quad}l}
\varepsilon_{\mathrm{bulk}}(\omega) & \mbox{inside tip}\\
 1 & \mbox{outside tip.}\end{array}\right.
\end{equation}
The resulting field then serves as input of the simulation of the electron emission. For the latter we employ three-dimensional quasi-classical trajectory Monte-Carlo simulations as well as one-dimensional time-dependent density functional theory (TDDFT) quantum simulations. Within TDDFT we treat the surface normal at the tip apex as the reaction coordinate. The tip radius $R\approx 6$ nm is large compared to the Fermi wavelength $\lambda_F\approx 4$ a.u.\ $\approx 0.2$ nm such that approximately translational symmetry in the surface plane can be assumed and at the same time small compared to the laser wavelength $\lambda=800$ nm such that the source field near the tip can be treated as homogeneous. The nano-scale dielectric response gives rise to a dramatic field enhancement (depending on tip parameters $F_{\mathrm{eff}}/F_0\approx 5$ to 10) and to a shift of the carrier-envelope phase $\Delta\phi_{\mathrm{CEP}}$ (Fig.\ \ref{fig2}).
For all tested configurations (tip radius, opening angle, material, pulse duration) no significant chirp or distortion of the envelope $f(t)$ was observed. The latter indicates that excitation of nanoplasmonic modes is of minor importance. In turn, the strong field enhancement leads to a significant reduction of $\gamma$ to an effective Keldysh parameter $\gamma_{\mathrm{eff}}$ thereby giving access to strong-field effects at moderate driving field strengths.

Within the adiabatic local-density approximation (LDA) of TDDFT \cite{Liebsch1997Electronic,Burke1998Density,Maitra2002Ten} the time-dependent electronic density $n(\vec r,t)$ is expanded in terms of one-body Kohn-Sham pseudo-wavefunctions $\psi_k(\vec r,t)$
\begin{equation}
n(\vec r,t)=\sum_{k=1}^{n_{\mathrm{occ}}} c_k |\psi_k(\vec r,t)|^2\, ,\label{eq2}
\end{equation}
where $n_{\mathrm{occ}}$ is the number of occupied orbits up to the Fermi energy. We use a metal slab of 200 a.u.\ width. The conduction band is represented by $n_{\mathrm{occ}}\sim 50$ orbitals. The weight coefficients $c_k$ are derived from the projection of the three-dimensional Fermi sphere onto the tip axis \cite{Eguiluz1984Static} such that Eq.\ \ref{eq2} gives the initial projected ground state density for $t\to -\infty$. Several ground state potentials including the self-consistent potential for a jellium slab and parameterized potentials with long-ranged image tails have been tested. The results presented in the following are only weakly dependent on their choice. The electron density is expressed in terms of the Wigner-Seitz radius $r_\mathrm{s}=2.334$ a.u.\ giving a Fermi energy of $E_\mathrm{F}=9.2$ eV. The work function of a clean tungsten (310) surface is $W_\mathrm{W(310)}=4.35$ eV. It is, however, sensitive to surface adsorbates \cite{Yamamoto1978} and can serve only as a first estimate. We have therefore checked on the work function dependence by varying $W$.

The time evolution of the electronic density is governed by the time-dependent Kohn-Sham equations
\begin{equation}
i \partial_t\psi_k(z,t) = \left\{-\frac{1}{2}\Delta + V[n(z,t)] + V_{\mathrm{ext}}(z,t)\right\}\psi_k(z,t)\, ,
\end{equation}
where $V[n(z,t)]$ contains the electrostatic and exchange-correlation potentials employing the LDA with the Wigner correlation functional. The potential of cores of the topmost atomic layer at which electrons rescatter is parameterized by a screened soft-core Coulomb potential
\begin{equation}
V_{\mathrm{atom}}(z)=- \frac{1}{1+|z|}  e^{-|z|/\lambda_{\mathrm{TF}}}
\end{equation}
with the Thomas-Fermi screening length $\lambda_{\mathrm{TF}}\approx 1$ a.u.\ for the electron gas. Our results are insensitive to the specific choice of $V_{\mathrm{atom}}$ as long as it is sufficiently strong to induce rescattering ($|V_{\mathrm{atom}}(z\approx 0)/2U_p| > 1$). The external potential is given by $V_{\mathrm{ext}}(z,t) = z F(t) + z F_{\mathrm{dc}}$. As in the experiment, a small static extraction field $F_{\mathrm{dc}}$ is included. The Kohn-Sham equations are integrated in real space by the Crank-Nicolson method with a constant time step of 0.05 a.u.\ over a total simulation time of 120 fs ($\sim 5000$ a.u.). The total size of the simulation box is 1425 a.u.\ (9500 grid points) with absorbing boundary conditions to avoid unphysical reflections due to the finite size of the system. Electron emission spectra are determined \cite{Pohl2000Towards} by a temporal Fourier transform of the wavefunctions at a detection point far from the surface ($\sim 900$ a.u.)  in order to ensure that the NIR field has terminated at the time of arrival of the wavepacket. Finally, the calculated spectra are broadened by $0.5$ eV to match the spectrometer resolution.

\begin{figure}
\centerline{\epsfig{file=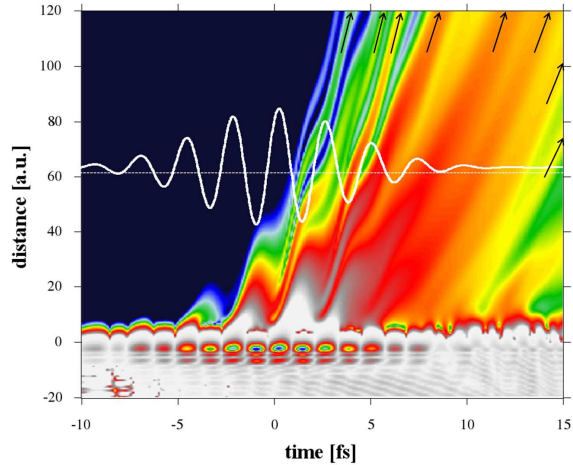,width=8cm}}
\caption{(Color online) Time-dependent change in electron density $|\delta n(z,t)| = |n(z,t)-n(z,-\infty)|$ on a logarithmic scale for a one-dimensional metal slab irradiated by a 6.5 fs laser pulse. The slope of equally colored lines signifies the momentum of emitted electrons. The color scale changes from logarithmic to linear at $z=0$ a.u. Electrons emitted near the field maxima (white solid line indicates $F+F_\mathrm{dc}$) are in part driven back to the surface leading to rescattering, wavepackets from subsequent cycles interfere (stripes visible for $z>50$ a.u., maxima indicated by arrows).}
\label{fig3}
\end{figure}

The time evolution of the NIR-field induced density fluctuations, $\delta n(z,t)=n(z,t)-n(z,-\infty)$ (Fig.\ \ref{fig3}), shows the onset of electron emission near the field maxima (e.g.\ near $t=-4.5$ fs).
After the initial acceleration towards vacuum, the electrons are driven back towards the surface after a change of sign of the laser field. Electrons (re-)scatter at the surface near the zero crossings of the electric field and are further accelerated giving rise to high kinetic energies in line with the simple man's model for atoms \cite{Corkum_3Step_1993}. Interference fringes clearly visible in Fig.\ \ref{fig3} as stripes at larger distances from the surface ($z\gtrsim 50$ a.u.) originate from successive emission events spaced by the laser period $T=2\pi/\omega$ giving rise to intercycle interferences or, equivalently, ATP peaks equispaced by $\hbar\omega$ in energy. A remarkable difference to atomic targets becomes apparent: sub-cycle (or intracycle) interferences resulting from electrons initially tunneling in opposite directions \cite{arbo2006} are absent due to the broken symmetry of the surface. The laser field inside the tip is effectively screened by an induced surface charge layer ($-10<z<0$ a.u., dark colored features in Fig.\ \ref{fig3}).

\begin{figure}
\centerline{\epsfig{file=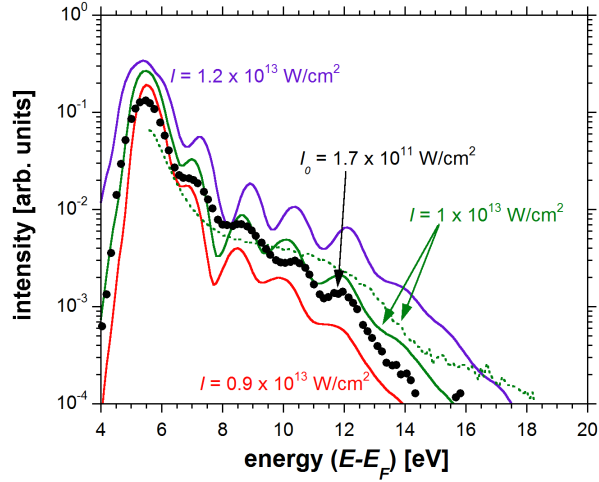,width=8cm}}
\caption{(Color online) Comparison of experimental and CEP averaged simulated spectra for a 6.5 fs laser pulse impinging on a tungsten tip. TDDFT results (solid lines) for different intensities are compared with an experimental spectrum (symbols) for $I_0=1.7\cdot 10^{11}$ W/cm$^2$ ($F_0\approx 2.2\cdot 10^{-3}$ a.u.) and a classical simulation (green dotted line) for an intensity of $I_{\mathrm{eff}}=10^{13}$ W/cm$^2$ ($F_0\approx 1.7\cdot 10^{-2}$ a.u.).}
\label{fig4}
\end{figure}

We complement our 1D-quantum simulations, which incorporate many-electron effects on the time-dependent mean-field level, by 3D-quasi-classical simulations on the single-active electron level in order to probe for effects due to the motion transverse to the laser polarization and tip axis neglected in the TDDFT calculations. The probability for an electron from the conduction band with kinetic energy $E_\perp$ perpendicular to the surface to tunnel through the surface barrier was taken to be proportional to $P(t)\propto \exp(-2\int dz\,\sqrt{2[V(z,t)-E_\perp]})$ weighted by the projection of the Fermi sphere onto the direction normal to the surface $\propto(E_\mathrm{F}-E_\perp)$. At the tunnel exit, the electrons acquire a randomly chosen $p_\perp$ normal to the surface, i.e.\ parallel to the laser polarization according to the ADK distribution \cite{DeloneKrainovMultiphoton1994}. For the momentum parallel to the surface $p_\|$ the width of the ADK distribution for atomic ionization can be considered to be the upper bound as the potential saddle at surfaces is broader leading to a broader distribution of the wavepacket in space and, consequently, a narrower distribution in momentum. We have therefore varied the width of the momentum distribution for $p_\|$ from $\sigma_{p_\|}=0$ to the width of the ADK distribution. Electrons returning to the surface are elastically scattered at atomic cores of the topmost layer. To simulate this process, doubly differential scattering cross sections have been calculated by a partial-wave analysis for scattering at a muffin-tin potential \cite{Christensen1974,Salvat2005ELSEPADirac}. For the energy range considered here the total cross section is larger than the size of a surface unit cell. Therefore, each electron is scattered off an atomic core upon return to the surface. Unlike in the case of atoms, the emitted electron does not have to return to near its parent atomic core but can also be backscattered from neighboring atoms. Taking an ADK width of the momentum component parallel to the surface of $\sigma_{p_\|}=0.1$ a.u.\ the wavepacket spreads upon rescattering over an area of more than 215 a.u.$^2$ covering $\sim 7.5$ surface unit cells on the W(310) surface. While the majority of electrons is scattered in forward direction, i.e.\ into the metal, a considerable fraction ($\lesssim 20$\%) is backscattered and is further accelerated by the laser field. The reason for the high intensity of the plateau is therefore twofold: the high density of scattering centers at the surface and the large large-angle scattering cross section for low-energy electrons. It is this anomalous enhancement of rescattering that also explains why a 1D quantum simulation, likely to overestimate rescattering for atomic targets, works surprisingly well for the surface of a nano-scale tip (Fig.\ \ref{fig4}).
Good agreement between CEP averaged experimental and simulated spectra is found for enhanced intensities ranging from $0.6$ to $1.5\cdot 10^{13}$ W/cm$^2$, an example of which is shown in Fig.\ \ref{fig4}. All spectra show similar features: a direct peak, a plateau only one to two orders of magnitude lower in intensity than the direct peak and an intensity-dependent cut-off energy. From comparison of the positions of the cut-off energies and the shape of experimental (orange spectrum of Fig.\ \ref{fig1}) and simulated spectra we conclude that the effective intensity is $I_{\mathrm{eff}}\approx 10^{13}$ W/cm$^2$ corresponding to a field enhancement of about 7.5. Surprisingly, the plateau area in the 3D classical simulation (green dotted line) is even more pronounced than in the experimental spectra which might be due to inelastic scattering events not included in this simulation.

Remarkably, best agreement with experiment is achieved when choosing a workfunction of $W\simeq 6.2$ eV in the TDDFT simulation. This is very close to the upper band edge of the tungsten $d$ electrons \cite{Christensen1974}. The role of $d$ electrons in ATP has been highlighted earlier \cite{Schenk2010}. However, in view of the influence of adsorption on experimental data and the simplifications underlying the TDDFT simulations, definite conclusions about the initial states of the photoelectrons are premature.

In conclusion, we have provided, both experimentally and theoretically, clear evidence of electron rescattering at tip-shaped metallic surfaces of nanometric dimensions. The signatures of strong-field physics, a plateau in the electron emission spectrum followed by a cut-off, appear at surprisingly low laser intensities of $I_0\lesssim 2 \cdot 10^{11}$ W/cm$^2$, well in the multi-photon regime. Dramatic field enhancement factors of $\sim 8$ leading to effective intensities $I_{\mathrm{eff}}\approx 10^{13}$ W/cm$^2$ near the tip are the core of the appearance of strong-field phenomena. Moreover, the high solid-state target density increases the overall probability for backscattering when the electron approaches the surface, thereby strongly enhancing the plateau heights. The shift in the CEP as predicted by the dielectric response may provide additional information on the collective electronic response in the metal not accessible with the present experimental setup. Future work will focus on the tip-induced shifts of $\phi_{\mathrm{CEP}}$ as a function of shape, material, and surface coverage of the metallic tip. Deeper analysis of data may also enable to extract quantities as scattering phase and electron dynamics on attosecond timescale.

This work was supported by the Austrian Science Foundation FWF under Proj.\ Nos.\ SFB-041 ViCoM and P21141-N16. M.S. and G.W. thank the International Max Plank Research School
on Advanced Photon Science for financial support. This work has been supported in part by the European Union (FP7-IRG).

\bibliography{nanotip}

\end{document}